\documentstyle[twocolumn,epsf]{jpsj}
\def\H{{\cal H}}
\def\v#1{\mib #1}
\def\ln{\mbox{ln}}
\def\JA{J_{\mbox{A}}}
\def\JB{J_{\mbox{B}}}

\def\runtitle{DMRG Study of the $S=1/2$ Quasiperiodic Heisenberg Antiferromagnets
}

\def\runauthor{Kazuo {\sc Hida}}

\title
{
Density Matrix Renormalization Group Study of the $S=1/2$ Antiferromagnetic Heisenberg Chains with Quasiperiodic Exchange Modulation}

\author
{Kazuo {\sc Hida}
\footnote{e-mail: hida@riron.ged.saitama-u.ac.jp}}

\inst
{
Department of Physics, Faculty of Science,\\ Saitama University, Urawa, Saitama 338-8570
}

\recdate
{}

\abst
{
The low energy behavior of the $S=1/2$ antiferromagnetic Heisenberg chains with precious mean quasiperiodic exchange modulation is studied by the density matrix renormalization group method. Based on the scaling behavior of the energy gap distribution, it is found that the ground state of this model belongs to the universality class different from that of the XY chain for which the precious mean exchange modulation is marginal. This result is consistent with the recent bosonization analysis of Vidal et al.\cite{vidal}}

\kword
{
Heisenberg model, precious mean, quasiperiodic, density matrix renormalization group, bosonization 
}
\begin{document}
\sloppy
\maketitle

In the recent studies of quantum many body problem, the low energy properties of the quantum spin systems with modulated spatial structure have been attracting a wide interest. Although the periodic chains and the random chains are studied in detail, the quasiperiodic chains, which has the intermediate character between the regular and random chains, are less studied except for the XY-case which is equivalent to the spinless free fermion chains\cite{kkt1,ko1,kst1,hk1,jh1}. In the fermionic language, the Ising component of the exchange coupling corresponds to the fermion-fermion interaction leading to the strong correlation effect which is the most important subject of the recent condensed matter physics. 

Although the XY chain can be mapped onto the free fermion chain, the problem is not trivial on the quasiperiodic lattice. For the Fibonacci lattice, Kohmoto and coworkers\cite{kkt1,ko1,kst1,hk1} clarified the Cantor set structure of the single particle spectrum and the wave function by means of the renormalization group (RG) method. Especially the dynamical exponents are found analytically at the band center and the band edge. Recently, this approach has been extended to include other types of quasiperiodic lattices and the anisotropy between $x$ and $y$ component of exchange couplings\cite{jh1}. It should be noted that the criticality of the Fibonacci XY chain stems from the marginal nature of the Fibonacci and other precious mean aperiodicity in this model. For the relevant aperiodicity, more singular behavior with divergent dynamical exponent is realized even for the XY chain\cite{jh1}. 

Although these works discovered the beautiful mathematical structure of quasiperiodic chains, almost no attempt to include the interaction effect is done so far except for the mean field approach\cite{hh1,hk1} and recent bosonization approach\cite{vidal}. In the present work, we employ the density matrix renormalization group (DMRG) method\cite{wh1,kh1} to take full account of the correlation effect in the $S=1/2$ precious mean antiferromagnetic Heisenberg chains which include the Fibonacci chains.

Our Hamiltionan is given by,
\begin{equation}
\label{eq:ham}
\H = \sum_{i=1}^{N-1} 2J_{\alpha_i}\v{S}_{i}\v{S}_{i+1},\ \ \ (J_{\alpha_i} > 0),
\end{equation}
where $\v{S}_{i}$'s are the spin 1/2 operators and the open boundary condition is assumed. The exchange couplings $J_{\alpha_i}$'s ($=\JA$ or $\JB$) follow the precious mean sequence generated by the substitution rule,
\begin{equation}
A \rightarrow A^k B, \ B \rightarrow A.
\end{equation}
The cases $k=1$ and $k=2$ correspond to the Fibonacci (golden mean) and silver mean chains, respectively. In the following, we take $\JA=1$ to fix the energy unit. For finite $N$, we consider all possible $(N-1)$-membered subsequences of the infinite precious mean chain and investigate the energy gap distribution among them. It should be noted that the number of the $(N-1)$-membered subsequence is equal to $N$\cite{penrose}.

In the XY case, the precious mean aperiodicity in the exchange coupling is marginal and the energy gap $\Delta$ scales with the system size as $\Delta \sim N^{-z}$ where $z$ is the dynamical exponent\cite{jh1}. Therefore the gap distribution function scales as 
\begin{equation}
\label{cr}
P(\Delta)d\Delta = N^z f(\Delta N^z) d\Delta.  
\end{equation}
Consequently, the average and flucutation of $\ln(1/\Delta)$ scales as,
\begin{eqnarray}
<\ln(1/\Delta)> &\simeq& C_1-z \ln N, \\ 
\sigma[\ln(\Delta)] &\equiv& \sqrt{<(\ln(\Delta)- <\ln(\Delta)>)^2>}, \nonumber \\
 &\simeq& \sqrt{C_2-C_1^2} = \mbox{const.}, 
\end{eqnarray}
where
\begin{displaymath}
C_n = \int_{-\infty}^{\infty}t^n g(t)dt, \ g(t)=f(\mbox{e}^t). 
\end{displaymath}
On the other hand, for the XY chain with relevant exchange aperiodicity, the gap scales as $\ln (1/\Delta) \sim N^{\omega}$\cite{jh1} and the gap distribution function scales as, 
\begin{equation}
\label{rs}
P(\ln\Delta)d\ln\Delta = N^{-\omega}f(N^{-\omega}\ln\Delta) d\ln\Delta,  
\end{equation}
which gives
\begin{eqnarray}
<\ln(1/\Delta)> &\simeq& D_1 N^{\omega}, \\
\sigma[\ln(\Delta)] &\simeq& \sqrt{D_2-D_1^2} N^{\omega},  
\end{eqnarray}
where
\begin{displaymath}
D_n = \int_{-\infty}^{\infty}x^n f(x)dx. 
\end{displaymath}
It should be remarked that $\sigma$ tends to a constant value for the marginal aperiodicity while it grows with the same exponent as $<\ln(1/\Delta)>$ for the relevant aperiodicity. This type of behavior with $\omega=1/2$ is observed also in the random singlet phase of the $S=1/2$ random antiferromagnetic Heisenberg chain.\cite{kh1,ds1}

The interacting spinless fermion chain with Fibonacci potential has been studied by Vidal et al.\cite{vidal} by means of the bosonization technique. This model can be mapped onto the XXZ chain in the Fibonacci magnetic field by the Jordan-Wigner transformation. In the following, we use the spin chain terminology. After bosonization, this model can be described by the boson Hamiltonian

\begin{equation} \label{Hbos}
H=H_0+H_W^h,
\end{equation}
\begin{equation} \label{H0bos}
H_0={1\over 2\pi}\int dx \left[(u K)(\pi \Pi)^2+\left({u\over K}\right)
(\partial_x \phi)^2\right],
\end{equation}
\begin{equation}
H_W^h = \frac1{2\pi \alpha} \int dx \,W(x)  \cos\left[2k_Fx+\sqrt{2}\phi(x)\right],
\label{hquasi}
\end{equation}
where $\phi$ is the boson field, $\Pi$, the field conjugate to $\phi$, $\alpha$, the ultraviolet cut-off, $u$, the spin wave velocity, $k_F$, the fermi wave number of the spinless fermions and $K$, the Luttinger liquid parameter. The function $W(x)$ represents the spatially varying magnetic field. The case $K=1$ corresponds to the $SU(2)$ invariant isotropic Heisenberg chain and $K=2$ to the XY chain. (Note that our definition of $K$ differs from that of ref. \citen{vidal} by a factor of 2.) For the Fibonacci type modulation, the function $W(x)$ is defined via its Fourier components given in ref. \citen{vidal}.

Using the standard bosonization scheme, the spatial modulation of the exchange coupling is similarly expressed as,
\begin{equation}
H_W^J = \frac1{2\pi \alpha} \int dx \, W(x)  \sin\left[2k_Fx+\sqrt{2}\phi(x)\right],
\end{equation}
which coincides with eq. (\ref{hquasi}) by the shift of the origin of $\phi$. Therefore the conclusion obtained by Vidal {\it et al.} also holds for the case of Fibonacci type exchange modulation given by the Hamiltonian (\ref{eq:ham}).

Vidal {\it et al.}\cite{vidal} derived the RG equation within the weak coupling approximation. Based on the numerical solution of the RG equation, they have shown that the Fibonacci modulation term becomes irrelevant or relevant according as $K > K_c$ or $K < K_c$ where $K_c \simeq 2$ in the absence of the uniform magnetic field.  For $K>K_c$, the ground state is the usual Luttinger liquid, while for $K < K_c$, the ground state is renormalized to the strong coupling regime and the weak coupling theory cannnot predict the ground states properties.  For $K=K_c$, which corresponds to the XY model, the Fibonacci modulation becomes marginal and the ground state is critical. This is consistent with the well-known case of the free spinless fermions in the Fibonacci potential. On the other hand, the isotropic case ($K=1$) is renormalized to the strong coupling regime.

\begin{figure}
\epsfxsize=70mm 
\centerline{\epsfbox{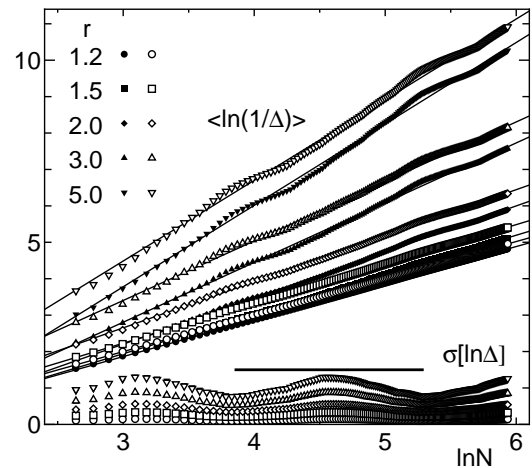}}
\caption{The $N$-dependence of $<\ln(1/\Delta)>$ and $\sigma[\ln\Delta]$  for the Fibonacci XY chain by the exact diagonalization method. The length of the horizontal bar is $3\ln((1+\sqrt{5})/2)$. In this and following figures \ref{fig3}, \ref{fig5} and \ref{fig6}, $r \equiv \mbox{Max} \{ \JB/\JA, \JA/\JB \}$  and the filled (open) symbols represent the case $\JB > \JA (\JB < \JA)$. The filled symbols for $\sigma$ almost overlap with open symbols.}
\label{fig1}
\end{figure}
\begin{figure}
\epsfxsize=70mm 
\centerline{\epsfbox{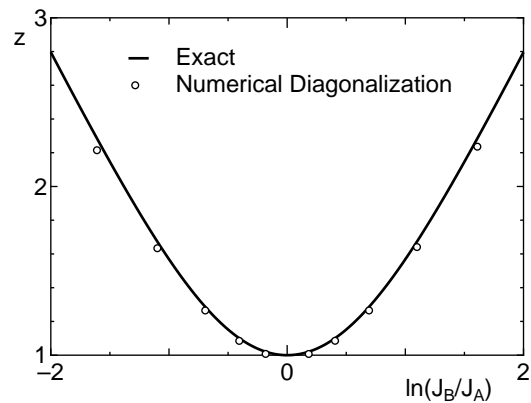}}
\caption{The dynamical exponent $z$ for the XY chain. The filled circle represent the present numerical calculation and solid line is the analytical results by Kohmoto {\it et al}.\cite{kkt1,kst1}}
\label{fig2}
\end{figure}

In the following, we  present the numerical results on the energy gap distribution. In the XY chain, the energy spectrum can be calculated numerically by the diagonalization of $N \times N$ matrices. The average $<\ln(1/\Delta)>$ and fluctuation $\sigma[\ln\Delta]$ of the logarithm of the energy gap for all possible $(N-1)$-membered subsequence is calculated for $14 \leq N \leq 378$ and various values of $\JB/\JA$ between 1/5 and 5. These are plotted against $\ln N$ in Fig. \ref{fig1}. The linearity of  $<\ln(1/\Delta)>$ to  $\ln N$ is fairly good and the flucutation $\sigma$ also tends to a constant value except for the oscillation with period $3 \ln\{(1+\sqrt{5})/2\}$ which is inherent to the Fibonacci chain spectrum\cite{luck,jh1}. The facter 3 comes from the fact that the single step of the RG transformation for the precious mean chains with odd $k$ corresponds to three inflation steps\cite{jh1}. We further calculated the dynamical exponent $z$ by fitting $\ln(1/\Delta)$-$\ln N$ curve by a straight line. The obtained values of $z$ are shown in Fig. \ref{fig2} by open circles. The solid line is the exact expression obtained by Kohmoto and coworkers\cite{ko1,kst1}. Again we find a good agreement. This result also confirms that the energy gap distribution of the finite length subsequences of the Fibonacci chain correctly reflects scaling properties of the ground state in the thermodynamic limit.
  
\begin{figure}
\epsfxsize=70mm 
\centerline{\epsfbox{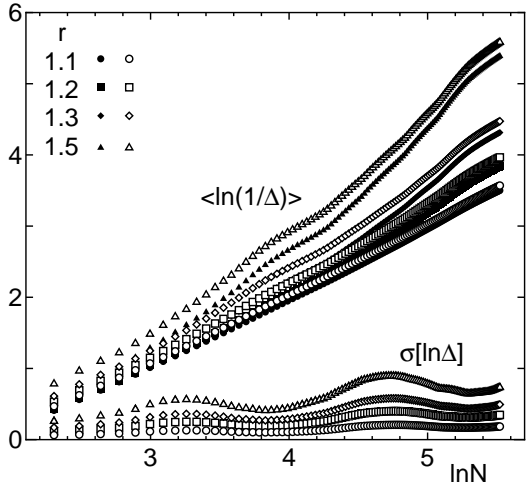}}
\caption{The $N$-dependence of $<\ln(1/\Delta)>$ and $\sigma[\ln\Delta]$ for the Fibonacci Heisenberg chain by the DMRG method plotted against $\ln N$.}
\label{fig3}
\end{figure}

The energy gap distribution for the Fibonacci Heisenberg chain is calculated by the DMRG method for  $14 \leq N \leq 234$ and various values of $\JB/\JA$ between 2/3 and 3/2 using the algorithm developed for the random chains\cite{kh1}. The number $m$ of the states kept at each step of DMRG iteration is 60 and only infinite size iterations are carried out. We have checked that $m$-dependence is negligible by increasing $m$ up to 80 for $J_B/J_A =3/2$ which is the most dangerous case studied here. For the further check of the accuracy of the DMRG scheme, we have also calculated the energy spectrum of the XY chain with $14 \leq N \leq 234$ using DMRG and found that the results coincide with the exact diagonalization results within the size of the symbols of Fig. \ref{fig1}.

\begin{figure}
\epsfxsize=70mm 
\centerline{\epsfbox{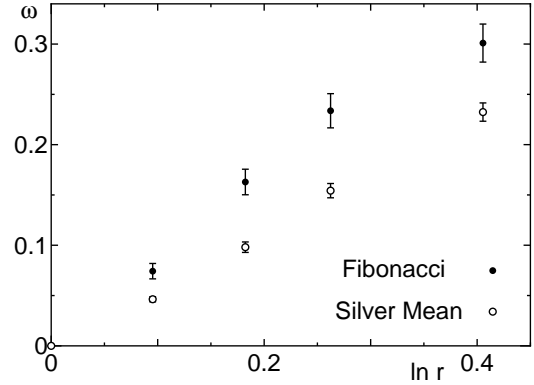}}
\caption{The exponent $\omega$ for the Fibonacci  and silver mean Heisenberg chains.}
\label{fig4}
\end{figure}

 Figure \ref{fig3} shows the average $<\ln(1/\Delta)>$ and fluctuation $\sigma[\Delta]$ plotted against $\ln N$ for the Fibonacci Heisenberg chain.  The curves of  $<\ln(1/\Delta)>$ show an evident upturn as $N$ increases. Further, the fluctuation $\sigma$ does not tend to a constant value. We have fitted the data for $<\ln(1/\Delta)>$ by the power law $N^{\omega}$. Such behavior is expected for the relevant aperiodicity in the XY case.\cite{jh1} The exponent $\omega$ turned out to be non-universal depending on the ratio $\JB/\JA$ as $\omega=\omega(\JB/\JA)$. We have further assumed that $\omega(\JB/\JA)=\omega(\JA/\JB)$ because the RG equation of ref. \citen{vidal} is invariant under the exchange $\JA \leftrightarrow \JB$. It should be also noted that the dynamical exponent for the precious mean XY chain is also invariant under the exchange $\JA \leftrightarrow \JB$. In addition, the numerically obtained values of $\omega$ for the silver mean chains also satisfiy the relation  $\omega(\JB/\JA) \simeq \omega(\JA/\JB)$ as explained below. So we assume that this relation also holds for the Fibonacci Heisenberg chains and use the average of the numerically obtained values of $\omega(\JB/\JA)$ and $\omega(\JA/\JB)$ as $\omega$. The values of $\omega$ are depicted against $\ln r$ ($r \equiv \mbox{Max}(\JB/\JA,\JA/\JB)$) in Fig. \ref{fig4} by filled symbols. The error bars in Fig. \ref{fig4} are estimated from the difference between $\omega(\JB/\JA)$ and $\omega(\JA/\JB)$. 

\begin{figure}
\epsfxsize=70mm 
\centerline{\epsfbox{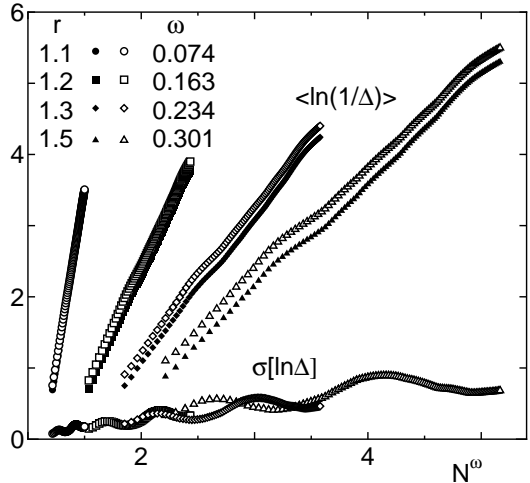}}
\caption{The $N$-dependence of $<\ln(1/\Delta)>$ and $\sigma[\ln\Delta]$ for the Fibonacci Heisenberg chain by the DMRG method plotted against $N^{\omega}$.}
\label{fig5}
\end{figure}

\begin{figure}
\epsfxsize=70mm 
\centerline{\epsfbox{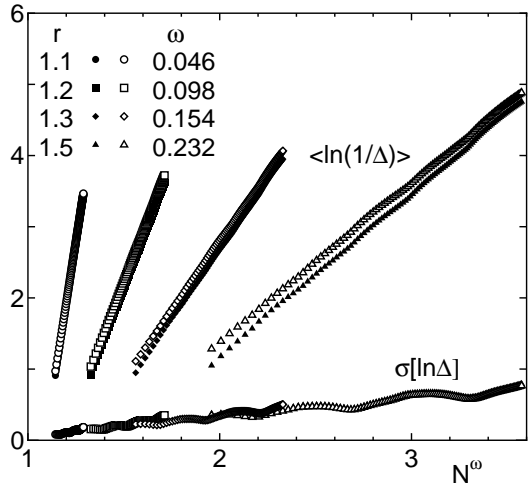}}
\caption{The $N$-dependence of $<\ln(1/\Delta)>$ and $\sigma[\ln\Delta]$ for the silver mean Heisenberg chain by the DMRG method plotted against $N^{\omega}$.}
\label{fig6}
\end{figure}

Using thus obtained values of $\omega$,  $<\ln(1/\Delta)>$ and $\sigma[\ln\Delta]$ are plotted against $N^{\omega}$ in Fig. \ref{fig5}.  It is clearly seen that both  $<\ln(1/\Delta)>$ and $\sigma[\ln\Delta]$ grow linearly with $N^{\omega}$. The same analysis is made for the silver mean chain in Fig. \ref{fig6} for $18 \leq N \leq 240$. In this case, the oscillation period in $\ln N$ is $\ln(1+\sqrt{2})$, because the single step of the RG transformation for the precious mean chains with even $k$ corresponds to a single inflation step\cite{jh1}. In this case, the numerically obtained values of $\omega$ satisfy $\omega(\JA/\JB)\simeq\omega(\JB/\JA)$ with better accuracy than the Fibonacci case. The values of $\omega$ are plotted against $\ln r$ in Fig. \ref{fig4} by open symbols. 

The above behavior of the energy gap clearly suggests that the ground state of the precious mean Heisenberg antiferromagnetic chain belong to the universality class different from that of the XY chain for which the precious mean aperiodicity is marginal. This result is consistent with the bosonization analysis of Vidal et al. \cite{vidal} Although these authors predict nothing about the ground state in the strong coupling regime, our numerical calculation shows that the gap distribution is  characterized by the scaling form (\ref{rs}) which is similar to that of the XY chain with relevant aperiodic modulation\cite{jh1}. It should be remarked in the random exhange case, the $S=1/2$ XXZ model ($1 \leq K \leq 2$) belong to the strong coupling regime because the critical value of $K$ is 3\cite{vidal,gia,doty}. For these models, the ground state is known to be the random singlet state which is also characterized by the gap distribution of the type (\ref{rs}) with $\omega=1/2$.\cite{ds1} Presumably, the gap distribution of  type (\ref{rs}) is the generic nature of the aperiodic exchange spin chains in the strong coupling regime. 

To summarize, based on the DMRG calculation of the gap distribution, it is found that the logarithm of the energy gap of the precious mean Heisenberg chains scales with a nonuniversal power of the system size as eq. (\ref{rs}). This is distinct from the critical gap distribution eq. (\ref{cr}) in the XY case. Taking the results for other kinds of aperiodic spin chains\cite{vidal,jh1,ds1,doty} into account, we expect that this is the characteristic property of the aperiodic spin chains in the strong coupling regime. 

In this paper, we concentrated on the energy gap distribution in the absence of the magnetic field. The effect of the magnetic field is interesting from two different points of view. First, in the presence of the uniform magnetic field, the multifractal Cantor-set structure of the single particle excitation spectrum of the free spinless fermion chain\cite{kst1,luck} manifests itself as the devil's staircase structure of the magnetization curve of the XY chain in the spin language. This can be regarded as the magnetization plateau problem\cite{khp,oya} in the quasiperiodic spatial structure. It is worth investigating if such structure survives in the Heisenberg case which has stronger quantum fluctuation than the XY case.

Another interesting problem is the effect of the precious mean modulation of magnetic field. For the random field XY chain, the band center state is known to be localized while it is the random singlet state with divergent spin correlation length for the random exchange XY chain although the randomness is relevant in both cases\cite{doty,mckenzie}. This difference comes from the perfect spin inversion symmetry of the random exchange problem\cite{doty}. From this point of view, effect of the precious mean modulation of the magnetic field might be different from that of the exchange coupling in the Heisenberg chain. This problem is left for future studies.

The author thanks  M. Kohmoto and T. Ohtsuki for valuable discussion and comments. The numerical calculations have been performed using the FACOM VPP500 at the Supercomputer Center, Institute for Solid State Physics, University of Tokyo and  the HITAC S820/80 at the Information Processing Center of Saitama University.  This work is supported by the Grant-in-Aid for Scientific Research from the Ministry of Education, Science, Sports and Culture.

\end{document}